# Realization of Quantum Spin Hall State in Monolayer 1T'-WTe$_2$

Shujie Tang[1, 2, 3, 4, 5†], Chaofan Zhang[1, 2†], Dillon Wong[6], Zahra Pedramrazi[6], Hsin-Zon Tsai[6], Chunjing Jia[1, 2], Brian Moritz[1], Martin Claassen[1], Hyejin Ryu[3, 7, 8], Salman Kahn[6], Juan Jiang[3, 9, 10], Hao Yan[1, 2], Makoto Hashimoto[11], Donghui Lu[11], Robert G. Moore[1, 2], Chancuk Hwang[9], Choongyu Hwang[8], Zahid Hussain[3], Yulin Chen[10], Miguel M. Ugeda[12,13], Zhi Liu[4,5], Xiaoming Xie[4,5], Thomas P. Devereaux[1, 2], Michael F. Crommie[6,14, 15], Sung-Kwan Mo[3*] & Zhi-Xun Shen[1, 2*]

[1]Stanford Institute for Materials and Energy Sciences, SLAC National Accelerator Laboratory, 2575 Sand Hill Road, Menlo Park, California 94025, USA

[2]Geballe Laboratory for Advanced Materials, Departments of Physics and Applied Physics, Stanford University, Stanford, California 94305, USA

[3]Advanced Light Source, Lawrence Berkeley National Laboratory, Berkeley, California 94720, USA

[4]State Key Laboratory of Functional Materials for Informatics, Shanghai Institute of Microsystem and Information Technology, Chinese Academy of Sciences, Shanghai 200050, China

[5]School of Physical Science and Technology, Shanghai Tech University, Shanghai 200031, China.

[6]Department of Physics, University of California at Berkeley, Berkeley, CA 94720, USA

[7]Max Plank POSTECH Center for Complex Phase Materials, POSTECH, Pohang 37673, Korea

[8]Department of Physics, Pusan National University, Busan 46241, Korea

[9]Pohang Accelerator Laboratory, POSTECH, Pohang 37673, Korea

[10]Department of Physics and Clarendon Laboratory, University of Oxford, Parks Road, Oxford, OX1 3PU, UK

[11]Stanford Synchrotron Radiation Lightsource, SLAC National Accelerator Laboratory, 2575 Sand Hill Road, Menlo Park, California 94025, USA

[12]Ikerbasque, Basque Foundation for Science, 48013 Bilbao, Spain

[13]CIC nanoGUNE, 20018 San Sebastián, Spain

[14]Materials Sciences Division, Lawrence Berkeley National Laboratory, Berkeley, California 94720, USA

[15]Kavli Energy Nano Sciences Institute, University of California and Lawrence Berkeley National Laboratory, Berkeley, California 94720, USA

†*These authors contributed equally to this work.*

*To whom correspondence should be addressed: skmo@lbl.gov, zxshen@stanford.edu*

**A quantum spin Hall (QSH) insulator is a novel two-dimensional quantum state of matter that features quantized Hall conductance in the absence of magnetic field, resulting from topologically protected dissipationless edge states that bridge the energy gap opened by band inversion and strong spin-orbit coupling. By investigating electronic structure of epitaxially grown monolayer 1T'-WTe$_2$ using angle-resolved photoemission (ARPES) and first principle calculations, we observe clear signatures of the topological band inversion and the band gap opening, which are the hallmarks of a QSH state. Scanning tunneling microscopy measurements further confirm the correct crystal structure and the existence of a bulk band gap, and provide evidence for a modified electronic structure near the edge that is consistent with the expectations for a QSH insulator. Our results establish monolayer 1T'-WTe$_2$ as a new class of QSH insulator with large band gap in a robust two-dimensional materials family of transition metal dichalcogenides (TMDCs).**

A two-dimensional (2D) topological insulator (TI), or a quantum spin Hall insulator, is characterized by an insulating bulk and a conductive helical edge state, in which carriers with different spins counter-propagate to realize a geometry-independent edge conductance $2e^2/h$ [1,2]. The only scattering channel for such helical edge current is back scattering, which is prohibited by time reversal symmetry, making QSH insulators a promising material candidate for spintronic and other applications.

The prediction of the QSH effect in HgTe quantum wells sparked the intense research efforts to realize the QSH state [3,4,5,6,7,8,9,10,11]. So far only a handful of QSH systems have been fabricated, mostly limited to quantum well structures of three-dimensional(3D) semiconductors such as HgTe/CdTe [3] and InAs/GaSb [6]. Edge conduction consistent with a QSH state has been observed [3,6,12]. However, the behavior under the magnetic field, where the time reversal symmetry is broken, cannot be explained within our current understanding of the QSH effect [13, 14]. There have been continued efforts to predict and investigate other material systems to further advance the understanding of this novel quantum phenomenon [5, 7, 8, 9, 15]. So far, it has been difficult to make a robust 2D material with QSH state, a platform needed for wide-spread study and application. The small band gaps exhibited by many candidate systems as well as their vulnerability to strain, chemical adsorption, and element substitution make them impractical for the advanced spectroscopic studies or applications. For example, a QSH insulator candidate stanene, a monolayer analog of graphene for tin, grown on $Bi_2Se_3$ becomes topologically trivial due to the modification of its band structure by the underlying substrate [11, 16]. The free standing Bi film with 2D bonding on a cleaved surface has shown edge conduction [9], but its topological nature is still debated [17]. It takes 3D out-of-plane bonding with the substrate and large strain (up to 18%) to open a bulk energy gap in monolayer bismuth [15]. Such 3D bonding structure may induce similar surface issues as seen in 3D semiconductor QSH systems. Monolayer FeSe grown on a $SrTiO_3$ substrate has also emerged as a model system to support both QSH and superconductivity. However, due to doping from the substrate the Fermi energy ($E_F$) is more than 500meV higher than the nontrivial gap, making it less practical for applications[18].

1T' phase monolayer TMDCs $MX_2$, M = (W, Mo) and X = (Te, Se, S), are theoretically predicted to be a promising new class of QSH insulators with large band gap [10]. Among them, $WTe_2$ is the only one for which the 1T' phase is most energetically favored. Realization of a QSH insulator in 2D TMDCs would be a breakthrough as this is a robust family of materials with none of the complications from surface/interface dangling bonds that are seen in 3D semiconductors, enabling a broad range of study and application of QSH physics. In this work, we report a successful growth of monolayer 1T'-$WTe_2$ using molecular beam epitaxy (MBE) on a bilayer graphene (BLG) substrate. In-situ ARPES measurements clearly show the band inversion and the opening of a 55meV bulk band gap, which is an order of magnitude larger than gaps seen in quantum wells of 3D semiconductors [3, 6]. Scanning tunneling spectroscopy (STS) spectra show evidence of the insulating bulk and conductive edge nature of 1T'-WTe$_2$. Our results thus provide compelling experimental evidences of a QSH insulator phase in monolayer 1T'-WTe$_2$.

Figure 1a presents the crystal structure of monolayer 1T'-WTe$_2$. MX$_2$ has three typical phases, namely 2H, 1T and 1T'. 1T-WTe$_2$ is composed of three hexagonally packed atomic layers in an ABC stacking. The metal atoms are in octahedral coordination with the chalcogen atoms. This is not a stable phase in freestanding form and undergoes a spontaneous lattice distortion into the 1T' phase via a doubling of the periodicity in the X direction. W atoms are dislocated from the original octahedral positions to form a zigzag chain in the Y direction.

The lattice distortion from the 1T phase to the 1T' phase induces band inversion and causes 1T'-WTe$_2$ to become topologically non-trivial [10, 19, 20]. Fig. 1b schematically summarizes this topological phase transition in 1T'-WTe$_2$. Without spin-orbit coupling (SOC), the inverted bands cross at a momentum point along the Γ-Y direction, forming a Dirac cone. Strong SOC lifts the degeneracy at the Dirac point, opening a bulk band gap. Following the bulk boundary correspondence [21, 22], the helical edge state is guaranteed by the gapped topologically non-trivial bulk band structure.

Our first-principles band structure calculations for 1T- and 1T'-WTe$_2$ are presented in Figs. 1c-e, which is generally consistent with the literature[10, 20, 23]. The key bands for the band inversion with opposite parities are marked to track their evolution. In 1T-WTe$_2$, the bands from *5d$_{xz}$* and *5d$_{z^2}$* orbitals of W are separated by the E$_F$ (Fig. 1c). Due to the symmetry breaking through the lattice distortion from 1T to 1T', these orbitals hybridize substantially. Fig. 1d shows that the $d_{z^2}$ orbital is lowered below $E_F$ whereas the $d_{xz}$ orbital lifts in the opposite direction near the Γ point. Because these two inverted bands have different parity at the Γ point, the $Z_2$ invariant $v$, in which $(-1)^v$ determined by the product of all occupied band parity eigenvalues [24], changes from 0 to 1. The valence band maximum in 1T' phase is mainly from W $d_{yz}$ orbital, with an even parity at the Γ point. When its degeneracy with $d_{xz}$ orbital in 1T phase is lifted by the lattice distortion, the band stays below $E_F$ and does not involve in the band inversion. With the inclusion of SOC (Fig. 1e), the bands further hybridize with each other and the degeneracies at the Dirac cones formed by the band inversion are lifted, opening a band gap in the bulk states. We note here that different calculation methods give different estimates on the size of the band gap for strain-free 1T'-WTe$_2$ monolayers. The generalized-gradient approximation (PBE) usually underestimates the bandgap and gives a negative band gap value[10], while PBE with hybrid function (HSE06) gives a positive value [23].

Figure 2 summarizes the MBE growth and the characterization of 1T'-WTe$_2$ on BLG/SiC(0001) (See Methods for the details of the growth condition). The reflection high-energy electron diffraction (RHEED) pattern of the BLG substrate and the monolayer 1T'-WTe$_2$ are presented in Fig. 2a. Clean vertical line profiles after the deposition of W and Te clearly indicate the layer-by-layer growth mode. Using the lattice constant of BLG (a=2.46 Å) as a reference, the lattice constant of the grown film is estimated to be ~ 6.3Å ± 0.2 Å, consistent with the expected value for monolayer 1T'-WTe$_2$ [23]. The angle-integrated core level photoemission spectrum (Fig. 2b) exhibits the characteristic peaks of W and Te for the 1T' phase. Two differently coordinated types of Te contribute two sets of Te *4d* peaks, while the clean doublet feature of the W *4f* peaks indicates a pure 1T' phase rather than a mixed phase of 1T' and 1H [25]. Fig. 2c is an atomically resolved scanning tunneling microscopy (STM) image of 1T'-WTe$_2$, from which a ~7.5° angle distortion is observed, which is universal in bulk 1T' phase $MX_2$ [25, 26]. The measured Fermi surface (FS) from the in-situ ARPES is shown in Fig. 2e. Due to the symmetry mismatch between the two-fold rotational symmetry of the sample and the three-fold symmetry of the substrate, there exist three energetically equivalent domains rotated by 120 degrees with respect to each other and each domain contributes two electron pockets along the ΓY direction of their respective BZs [27]. The experimental band dispersion along ΓY cutting the FS electron pockets is inevitably superposed with the contributions from ΓP and ΓP'. However, as shown in Fig. 2f - h, the valence bands from ΓP' and ΓP directions are enclosed by the ΓY band. Therefore, the existence of the multiple domains does not affect the characterization of the gap size and the separation between valence and conduction bands. Overall band structure measured with ARPES (Figs. 2f & g) gives a nice agreement with the HSE06 calculation (Fig. 2g), demonstrating the 1T' nature and the high quality of our thin film samples. The predicted band inversion in 1T'-$WTe_2$ is well established experimentally by a polarization-dependent ARPES measurement, from which one can clearly distinguish in- and out-of-plane orbital characters and their inversion around the Γ point (Supplementary Figs. S1 and S2). This indicates the nontrivial topology of the 1T'-$WTe_2$.

The signature of strong SOC in 1T'-$WTe_2$ is the lifting of state degeneracy at the Dirac cones along the ΓY direction, resulting in an opening of the bulk gap as illustrated in Fig. 3a. This can be seen more clearly in the energy distribution curves (EDCs) extracted at the valence band top and the conduction band bottom. Since ARPES data in Fig. 2 only show faint tails of the bulk conduction band, we deposited potassium (K) onto the surface to raise $E_F$ [28] and make the conduction band more clearly visible to ARPES. Fig. 3b focuses only the low energy electronic structure of surface K-doped 1T'-$WTe_2$, with $E_F$ raised ~ 70meV to reveal the conduction band bottom more clearly. The corresponding EDCs in Fig. 3c show that the conduction band and the valence band are well separated from each other. To quantify the size of the band gap, we extracted two EDCs from the momentum positions at the conduction band bottom and valence band top, labeled by the dashed lines in Fig. 3b, and we overlaid them in Fig. 3d. The red and green peaks in Fig. 3d corresponds to the energy positions of the conduction band bottom and the valence band top, respectively. We estimate the size of the band gap to be 55 ± 20meV and 45 ± 20meV in intrinsic and K-doped samples, respectively (Supplementary Fig. S3). This is in clear contrast to the bulk 1T'-WTe$_2$, which is a semimetal with a complex band structure near E$_F$ exhibiting multiple Fermi pockets [29]. The stacking of energy momentum dispersions with fine momentum steps parallel to the ΓY direction (Figs. 3e and 3f) further establish the effect of SOC by showing that the gap never closes for any momentum across the FS.

Now that we have established the band inversion and the opening of a band gap due to the strong SOC, the remaining signature of a QSH insulator is the conductive edge state in contrast to the insulating bulk, which can be better examined by STS. Fig. 4a shows the local differential conductance (dI/dV) spectrum taken at a point far away from the WTe$_2$ edges, which represents the bulk local density of states (LDOS). The peak positions in dI/dV are in good agreement with the band edges found in ARPES. The agreement between ARPES and STS further extends to the size of the gap, as the mean gap size determined by STS is 56 ± 14 meV (Supplementary Figs. S6). In contrast to the gap in the bulk, dI/dV at a 1T'-WTe$_2$ edge is quite different, showing a "V-shape" spectrum with states filling in the bulk gap (Fig. 4b), which may indicate the existence of a conductive edge state. Indeed, similar dI/dV spectral line shapes have been reported for other topological systems with distinct edge states [15, 30] and have been attributed to the one-dimensional (1D) nature of the edge states and the emergence of a Luttinger liquid [15]. Fig. 4c shows dI/dV as a function of energy and distance away from an edge, which demonstrates that the V-shaped conductance is localized at the edge of the WTe$_2$. We observe that such localized edge states run continuously along our sample edges (Supplementary Figs. S7), with only small variations in the fine details of the spectra, regardless of the size, shape, and edge-roughness of samples. This provides evidence of the edge state's topologically non-trivial nature. [30, 31]

By combining ARPES and STS results, we provide strong evidence supporting the direct observation of all the characteristic electronic properties of the a QSH state with large energy gap in 1T'-WTe$_2$, confirming the theoretical prediction [10]. Such a robust platform for QSH insulator in 2D TMDCs should provide new opportunities for fundamental studies and novel device applications. Since TMDCs are inert, widely available, can be exfoliated for transport experiments, and be made into few-layer and van der Waals heterostructure devices, we expect them to be the material of choice for much expanded, multimodal effort to understand and utilize QSH systems.

## References

1. Kane CL, Mele EJ. Quantum Spin Hall Effect in Graphene. *Physical Review Letters* 2005, **95**(22)**:** 226801.

2. Bernevig BA, Hughes TL, Zhang S-C. Quantum Spin Hall Effect and Topological Phase Transition in HgTe Quantum Wells. *Science* 2006, **314**(5806)**:** 1757-1761.

3. König M, Wiedmann S, Brüne C, Roth A, Buhmann H, Molenkamp LW*, et al.* Quantum Spin Hall Insulator State in HgTe Quantum Wells. *Science* 2007, **318**(5851)**:** 766-770.

4. Liu C, Hughes TL, Qi X-L, Wang K, Zhang S-C. Quantum Spin Hall Effect in Inverted Type-II Semiconductors. *Physical Review Letters* 2008, **100**(23)**:** 236601.

5. Liu Z, Liu C-X, Wu Y-S, Duan W-H, Liu F, Wu J. Stable Nontrivial $Z_2$ Topology in Ultrathin Bi (111) Films: A First-Principles Study. *Physical Review Letters* 2011, **107**(13)**:** 136805.

6. Knez I, Du R-R, Sullivan G. Evidence for Helical Edge Modes in Inverted InAs/GaSb Quantum Wells. *Physical Review Letters* 2011, **107**(13)**:** 136603.

7. Xu Y, Yan B, Zhang H-J, Wang J, Xu G, Tang P*, et al.* Large-Gap Quantum Spin Hall Insulators in Tin Films. *Physical Review Letters* 2013, **111**(13)**:** 136804.

8. Weng H, Dai X, Fang Z. Transition-Metal Pentatelluride $ZrTe_5$ and $HfTe_5$: A Paradigm for Large-Gap Quantum Spin Hall Insulators. *Physical Review X* 2014, **4**(1)**:** 011002.

9. Drozdov IK, Alexandradinata A, Jeon S, Nadj-Perge S, Ji H, Cava RJ*, et al.* One-dimensional topological edge states of bismuth bilayers. *Nat Phys* 2014, **10**(9)**:** 664-669.

10. Qian X, Liu J, Fu L, Li J. Quantum spin Hall effect in two-dimensional transition metal dichalcogenides. *Science* 2014, **346**(6215)**:** 1344-1347.

11. Zhu F-f, Chen W-j, Xu Y, Gao C-l, Guan D-d, Liu C-h*, et al.* Epitaxial growth of two-dimensional stanene. *Nat Mater* 2015, **14**(10)**:** 1020-1025.

12. Nowack KC, Spanton EM, Baenninger M, König M, Kirtley JR, Kalisky B*, et al.* Imaging currents in HgTe quantum wells in the quantum spin Hall regime. *Nat Mater* 2013, **12**(9)**:** 787-791.

13. Ma EY, Calvo MR, Wang J, Lian B, Muhlbauer M, Brune C*, et al.* Unexpected edge conduction in mercury telluride quantum wells under broken time-reversal symmetry. *Nat Commun* 2015, **6:** 7252.

14. Fabrizio N, Henri JS, Morten K, Charles MM, Ebrahim S, Joshua AF*, et al.* Edge transport in the trivial phase of InAs/GaSb. *New Journal of Physics* 2016, **18**(8)**:** 083005.

15. Reis F, Li G, Dudy L, Bauernfeind M, Glass S, Hanke W*, et al.* Bismuthene on a SiC Substrate: A Candidate for a New High-Temperature Quantum Spin Hall Paradigm. Arxiv; 2016.

16. Xu Y, Tang P, Zhang S-C. Large-gap quantum spin Hall states in decorated stanene grown on a substrate. *Physical Review B* 2015, **92**(8)**:** 081112.

17. Takayama A, Sato T, Souma S, Oguchi T, Takahashi T. One-Dimensional Edge States with Giant Spin Splitting in a Bismuth Thin Film. *Physical Review Letters* 2015, **114**(6)**:** 066402.

18. Wang ZF, Zhang H, Liu D, Liu C, Tang C, Song C*, et al.* Topological edge states in a high-temperature superconductor FeSe/SrTiO3(001) film. *Nat Mater* 2016, **15**(9)**:** 968-973.

19. Keum DH, Cho S, Kim JH, Choe D-H, Sung H-J, Kan M*, et al.* Bandgap opening in few-layered monoclinic MoTe$_2$. *Nat Phys* 2015, **11**(6)**:** 482-486.

20. Choe D-H, Sung H-J, Chang KJ. Understanding topological phase transition in monolayer transition metal dichalcogenides. *Physical Review B* 2016, **93**(12)**:** 125109.

21. Essin AM, Gurarie V. Bulk-boundary correspondence of topological insulators from their respective Green's functions. *Physical Review B* 2011, **84**(12)**:** 125132.

22. Hasan MZ, Kane CL. *Colloquium*: Topological insulators. *Reviews of Modern Physics* 2010, **82**(4)**:** 3045-3067.

23. Zheng F, Cai C, Ge S, Zhang X, Liu X, Lu H*, et al.* On the Quantum Spin Hall Gap of Monolayer 1T'-WTe$_2$. *Advanced Materials* 2016, **28**(24)**:** 4845-4851.

24. Fu L, Kane CL. Topological insulators with inversion symmetry. *Physical Review B* 2007, **76**(4)**:** 045302.

25. Das PK, Di Sante D, Vobornik I, Fujii J, Okuda T, Bruyer E*, et al.* Layer-dependent quantum cooperation of electron and hole states in the anomalous semimetal WTe$_2$. *Nat Commun* 2016, **7:** 10847.

26. Hla SW, Marinković V, Prodan A, Muševič I. Proceedings of the 15th European Conference on Surface ScienceSTM/AFM investigations of β-MoTe$_2$, α-MoTe$_2$ and WTe$_2$. *Surface Science* 1996, **352:** 105-111.

27. Ugeda MM, Bradley AJ, Zhang Y, Onishi S, Chen Y, Ruan W*, et al.* Characterization of collective ground states in single-layer NbSe$_2$. *Nat Phys* 2016, **12**(1)**:** 92-97.

28. Zhang Y, Chang T-R, Zhou B, Cui Y-T, Yan H, Liu Z, *et al.* Direct observation of the transition from indirect to direct bandgap in atomically thin epitaxial MoSe$_2$. *Nat Nano* 2014, **9**(2)**:** 111-115.

29. Jiang J, Tang F, Pan XC, Liu HM, Niu XH, Wang YX, *et al.* Signature of Strong Spin-Orbital Coupling in the Large Nonsaturating Magnetoresistance Material WTe$_2$. *Physical Review Letters* 2015, **115**(16)**:** 166601.

30. Pauly C, Rasche B, Koepernik K, Liebmann M, Pratzer M, Richter M, *et al.* Subnanometre-wide electron channels protected by topology. *Nat Phys* 2015, **11**(4)**:** 338-343.

31. Lin H, Das T, Okada Y, Boyer MC, Wise WD, Tomasik M, *et al.* Topological Dangling Bonds with Large Spin Splitting and Enhanced Spin Polarization on the Surfaces of Bi2Se3. *Nano Letters* 2013, **13**(5)**:** 1915-1919.

## Acknowledgments

The ARPES and thin film growth works at the Stanford Institute for Materials and Energy Sciences and Stanford University are supported by the Office of Basic Energy Sciences, Division of Materials Science and AFOSR Grant FA9550-14-1-0277 through the University of Washington. Research performed at ALS (thin film growth and ARPES) is supported by the Office of Basic Energy Sciences, US DOE under Contract No. DE-AC02-05CH11231. SSRL is supported by the Office of Basic Energy Sciences, US DOE under Contract No. DE-AC02-76SF00515. Scanned probe measurements were supported by the VdW Heterostructure program (KCWF16) (STM spectroscopy) funded by the Office of Basic Energy Sciences, US DOE under Contract No. DE-AC02-05CH11231, as well as by National Science Foundation award EFMA-1542741 (surface treatment and topographic characterization). Z. L. is supported by the National Natural Science Foundation of China (11227902). S. T. acknowledges the support by CPSF-CAS Joint Foundation for Excellent Postdoctoral Fellows. J. J. and C. C. H. acknowledge the support of the NRF, Korea through the SRC center for Topological Matter (No. 2011-0030787). H. R. and C. H. acknowledge the support from the NRF, Korea through Max Planck Korea/POSTECH Research Initiative (No. 2011-0031558) and Basic Science Research Program (No. 2015R1C1A1A01053065). A portion of the computational work was performed using the resources of the National Energy Research Scientific Computing Center (NERSC) supported by the U.S. Department of Energy, Office of Science, under Contract No. DE-AC02-05CH11231.

## Author contributions

S. T., C. Z., S. K. M. and Z. X. S. proposed and designed the research. S. T. performed the MBE growth. S. T. and C. Z. carried out the ARPES measurements and analyzed the ARPES data with help from J. J., H. R., S. K. M., M. H., D. L..  D. W., Z. P., H. T., S. K., M. M. U. and M. F. C. performed the STM measurements and analyzed STM data. C. J., M. C., B. M. and T. P. D. carried out the density functional calculations and provided theoretical support. S. T., S. K. M., D. W. and Z. X. S. wrote the manuscript with contributions and comments from all authors.

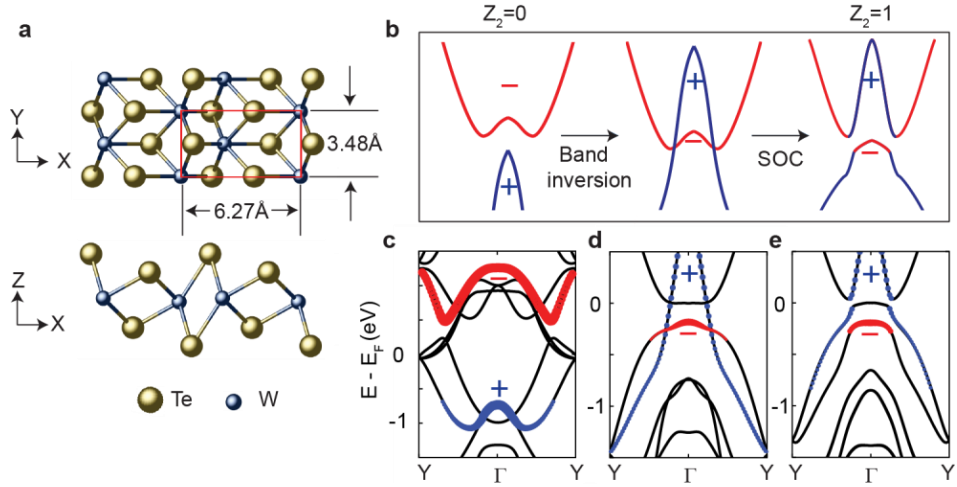

**Fig. 1**. **Topological phase transition in 1T'-WTe$_2$.** (a) Crystal structure of 1T'-WTe$_2$. The doubled period due to the spontaneous lattice distortion from 1T phase is indicated by the red rectangle. (b) Schematic diagram to show the bulk band evolution from a topologically trivial phase, to a non-trivial phase, and then to a bulk band opening due to SOC. Calculated band structures for WTe$_2$ to show the evolution from (c) 1T-WTe$_2$ along ΓY direction, (d) 1T'-WTe$_2$ without SOC, and (e) 1T'-WTe$_2$ with SOC. Red and blue dotted bands highlight the two bands involved in band inversion, which mainly contain the $5d_{z^2}$ and $5d_{xz}$ orbital contents, respectively. + and - signs denote the parity of the Bloch states at the Γ point.

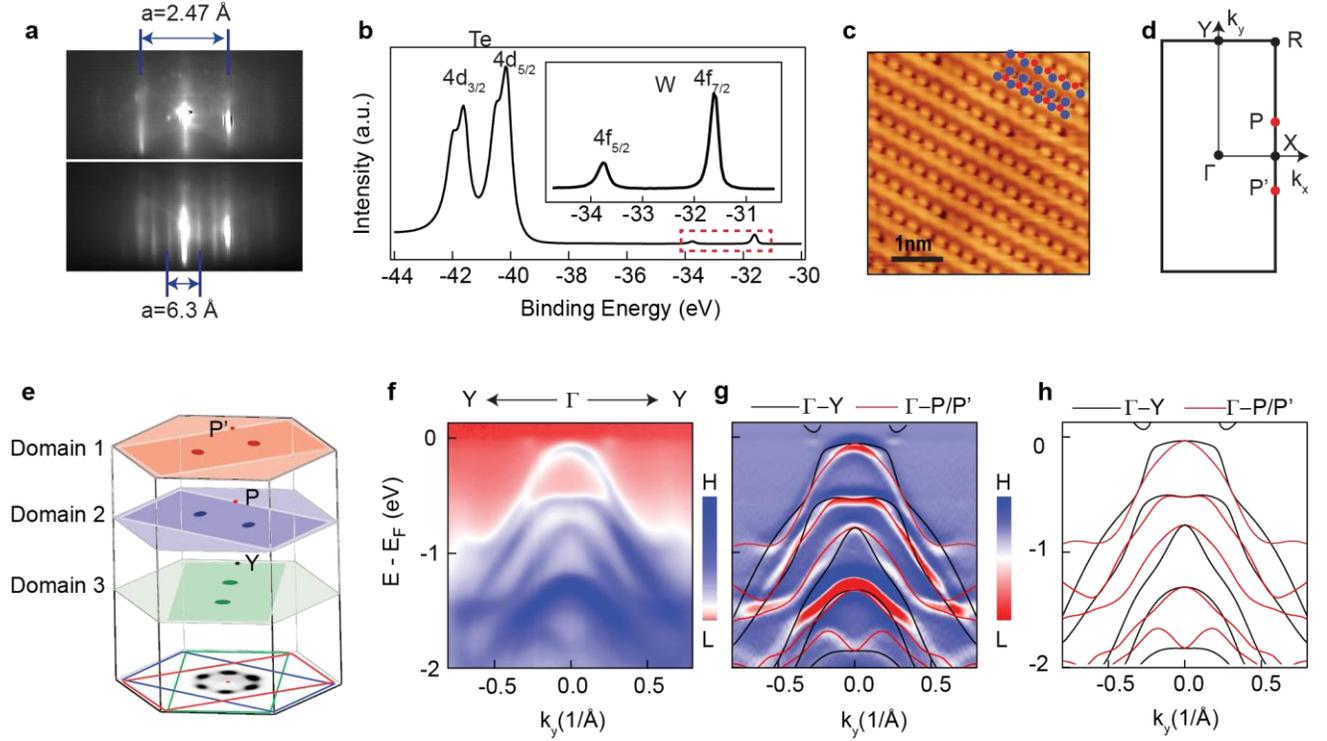

**Fig. 2**. **Characterization of epitaxially grown 1T'-WTe$_2$ and overall electronic structure from ARPES.** (a) RHEED pattern of graphene substrate (top) and sub-monolayer 1T'-WTe$_2$ (bottom). (b) Core level spectra of 1T'-WTe$_2$. The inset is a close-up for the region marked by red dashed rectangle. (c) Atomically-resolved STM topographic image of 1T'-WTe$_2$. Blue and red dots represent W and Te atoms, respectively. (d) Brillouin zone of 1T'-WTe$_2$. Time reversal invariant momenta Γ, X, Y, R are labeled by black dots. (e) Fermi surface map of 1T'-WTe$_2$. The intensity is integrated within a ±10 meV window around E$_F$. There are three domains rotated with respect to each other by 120 degrees due to difference in the symmetry of the sample and the substrate. The measured data along the ΓY high symmetry direction is unavoidably mixed with the signals from the ΓP and ΓP' directions. The schematic contributions from different domains are represented by different color panes above the real Fermi surface map at the bottom.

(f) Overall band structure measured along the experimental ΓY direction. (g) The second derivative spectra to enhance low intensity features. The overlaid black lines are the calculated band structure along the ΓY direction. (h) Calculated band structure along the ΓY (black) and ΓP/P' (red) directions, respectively. The low energy electronic structure around the Γ-point is dominated by the contributions from the ΓY bands.

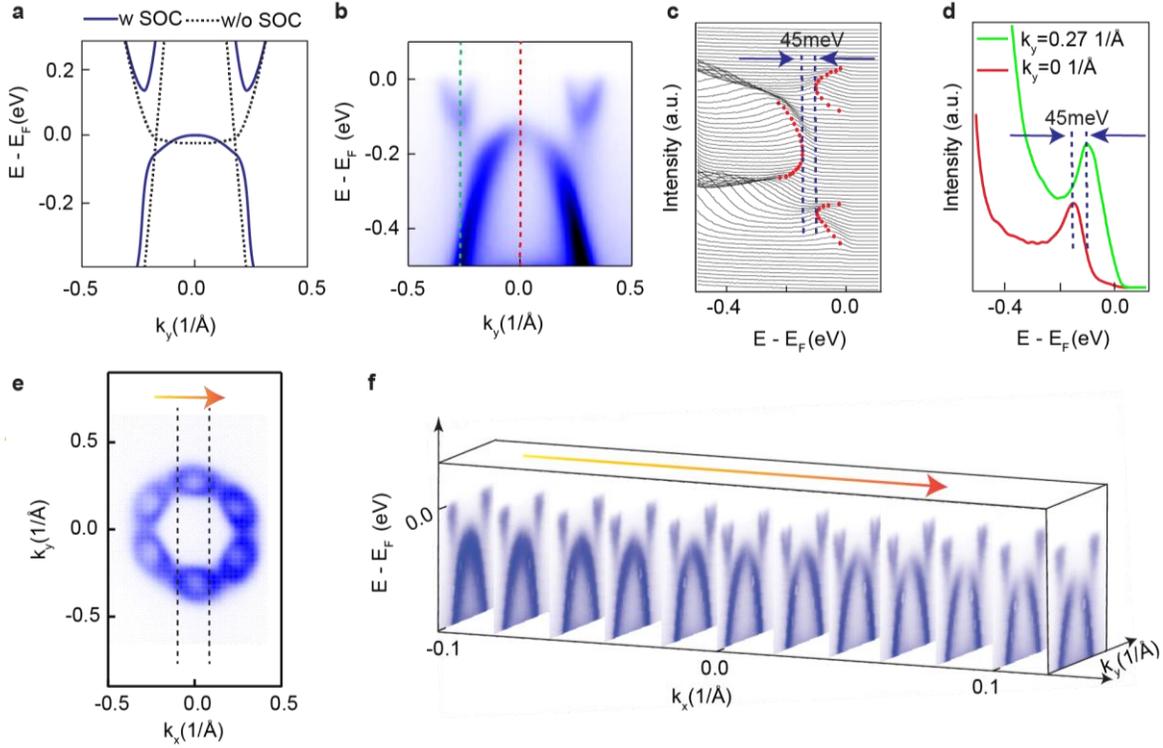

**Fig. 3**. **Band gap opening in monolayer 1T'-WTe₂** (a) Calculated band structure along the ΓY direction. (b) ARPES data along ΓY direction taken from surface K-doped sample. (c) EDCs for the data in panel (b). (d) EDCs from the momentum positions marked with green and red lines in (b). The green line corresponds to the conduction band bottom and the red line corresponds to the valence band top. (e) Fermi surface map of K-doped sample. Six electron pockets are due to the 3 rotational domains as explained for Fig. 2e. We only focus on the FS from a single domain. (f) Stacking plot of cuts between the parallel dotted lines labeled in (g).

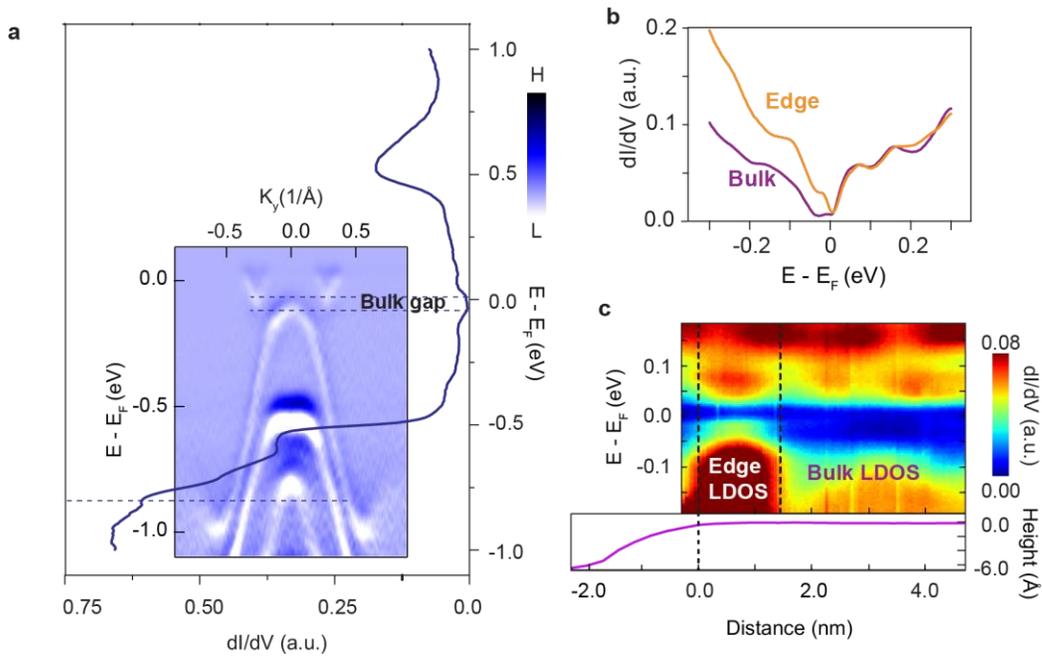

**Fig. 4**. **Tunneling spectroscopy in the bulk and at the edge of 1T'-WTe₂.** (a) STM dI/dV spectrum acquired in the bulk of monolayer 1T'-WTe₂. The inset is the high symmetry ARPES cut along the ΓY direction aligned in energy with the STS spectrum (acquired from a K-doped sample). Since the surface K-doping raises the position of $E_F$ by 70meV, the whole ARPES spectrum is shifted by that amount for proper comparison with STS. (b) Representative dI/dV spectra taken at the edge (orange) and in the bulk (purple), respectively. (c) dI/dV spectra taken across the step edge of a 1T'-WTe₂ monolayer island (top), and corresponding height profile (bottom).

# Supplementary Information for

**Realization of Quantum Spin Hall State in Monolayer 1T'-WTe$_2$**

Shujie Tang, Chaofan Zhang, Dillon Wong, Zahra Pedramrazi, Hsin-Zon Tsai, Chunjing Jia, Brian Moritz, Martin Claassen, Hyejin Ryu, Salman Kahn, Juan Jiang, Hao Yan, Makoto Hashimoto, Donghui Lu, Robert G. Moore, Chancuk Hwang, Choongyu Hwang, Zahid Hussain, Yulin Chen, Miguel M. Ugeda, Zhi Liu, Xiaoming Xie, Thomas P. Devereaux, Michael F. Crommie, Sung-Kwan Mo* & Zhi-Xun Shen*

correspondence to: skmo@lbl.gov, zxshen@stanford.edu

.

**Polarization dependent ARPES**

A powerful and direct way to discern the orbital characters of particular bands is the polarization dependent ARPES measurements. The matrix element of a photoemission process can be written as $\left|M_{f,i}^{k}\right|^{2} \propto \left|\langle\phi_{f}^{k}|\hat{\varepsilon}\cdot\boldsymbol{r}|\phi_{i}^{k}\rangle\right|^{2}$, where $\hat{\varepsilon}$ is the unit vector of the electric field of the light, $\phi_{f}^{k}$ and $\phi_{i}^{k}$ are the initial and final state wave functions of the photoelectron, and $\boldsymbol{r}$ is the position of electrons. By changing $\hat{\varepsilon}$ with respect to other geometric factors, one may enhance or suppress the ARPES signal from the orbitals of particular symmetry. Within our experimental configuration sketched in Fig. S1a, both the analyzer slit and incident light are in the mirror plane defined by the analyzer slit and the sample surface normal. In such a setup, P and S polarized photons make $\hat{\varepsilon}\cdot\boldsymbol{r}$ even and odd, respectively. With the final state of photoelectron approximated as an even parity plane-wave state $e^{i\boldsymbol{k}\cdot\boldsymbol{r}}$ with respect to the mirror plane, the photoemission signal from an initial state with even (odd) parity with respect to mirror plan are always enhanced (depressed) under P polarized photon excitation. The reverse is true for S polarized incident light. The measured polarization dependence can be directly compared to the orbital characters from the band calculation (Fig. S1b), which is obtained from Ref. 20.

Figures. S1c and d are ARPES data along the ΓY direction with P and S polarized incident light. To see the low energy electronic structure including the conduction band more clearly, we performed potassium (K) surface doping on the sample, which is known to inject electrons to a sample and raise chemical potential. Figs. S1e and f only focus on the low energy band structure along ΓY from the surface K-doped samples with P and S polarizations, respectively. To guide the eye, we marked three segments on different bands by red dots in Fig. S1c and e. It is clear that the red dotted segments show high intensity only with P polarization but can hardly be seen with S polarization. By rotating the azimuthal angle of the sample, one can change the orbital reflected on the mirror plane and get more information on the orbital characters of the bands. We focus on the predicted inverted band, the second valence band from the Fermi energy. Different ARPES data are taken with the different azimuthal angle, 0, 29 and 90 degrees in Fig. S2. The intensity of second valence band is always suppressed at S-polarization and enhanced at P-polarization. This indicates the symmetry of this band is always even with respect to the mirror plane we chose. We thus conclude that the only orbital satisfying this criterion is the $d_{z^2}$ orbital, which agrees well with the calculation (red dotted band in Fig. S1b). At the same time, intensity from the orbitals with in-plane character is clearly enhanced for S polarization as can be seen in Figs. S1d and f. Our polarization dependent ARPES gives strong evidences that band inversion occurs in our 1T'-WTe$_2$ monolayer films.

**Fitting the EDC peaks**

In order to extract the exact bulk band gap size, one needs to get the peak position of conduction band bottom and valence band top. We used Gaussian peaks with Shirley background to fit the energy distribution curves (EDCs). The fitted peaks are black dotted lines in Fig. S3, which fit the original EDCs very well. Each Gaussian peak of the fit is plotted as a blue solid line, and the corresponding peak position is labeled above it. The EDCs at k=0 1/Å and k=0.27 1/Å correspond to the valence band top and conduction band bottom, respectively. The gap sizes determined from the peak-to-peak distance are 55 ± 20 meV and 45 ± 20 meV for intrinsic and K-doped samples, respectively. The decrease of the gap size after surface doping is expected due to the surface doping inducing a vertical electric field without changing the non-trivial topology of 1T'-WTe$_2$[10].

**Scanning tunneling microscopy and gap determination**

To protect the samples from air during transit to the STM chamber, the 1T'-WTe$_2$ samples were capped with Se and Te layers. To acquire STM and STS data on the 1T'-WTe$_2$, the capping layers must be removed by annealing in UHV. Fig. S4a shows a topographic image of 1T'-WTe$_2$ domains after annealing the sample to T = 200º C. Most WTe$_2$ islands are roughly 20 nm x 20 nm in size, although there is some variation depending on growth conditions. The height of 1T'-WTe$_2$ above BLG is 9 ± 1 Å, but the 1T'-WTe$_2$ islands are usually surrounded (or partially surrounded) by a 1~3 Å high contaminant on the BLG (likely unreacted W or remnants of the Te capping layer). The contaminant can be removed from BLG by annealing to T = 350º C, but this induces disorder in most of the 1T'-WTe$_2$ edges. The contaminant can also be moved with the STM tip, and its presence or absence has no discernable effect on STS results for 1T'-WTe$_2$.

Two control experiments to evaluate the effect of the capping-decapping process on the electronic structure of 1T'-WTe$_2$ were performed. (1) After our in situ ARPES measurements, we capped the sample in the same manner as described above. Then the capping layer was removed afterwards, followed by ARPES measurements of the surface that went through the capping/decapping process. No difference was found in the ARPES spectra between the fresh sample surface and the surface that went through the capping-decapping process. (2) Besides using a Te/Se double layer as the capping layer (Te in direct contact with the film), a Se-only capping layer (no Te) was also used to protect the sample during the transfer between STM and MBE chamber. No noticeable difference was found from STM measurements on these two types of samples. Therefore, we believe that the capping-decapping process does not alter the measured electronic structure.

Figure S5c shows a representative 2 nm x 10 nm topographic image of a 1T'-WTe$_2$ edge. Horizontal stripes separated by ~6 Å can be seen on the 1T'-WTe$_2$. The slope of the step edge is due to the finite size of the STM tip.

Figure S5d shows the Fourier transform of the atomically resolved topographic image in Fig. 2c of the main text, from which we can measure the 1T'-WTe$_2$ lattice parameters to be 3.6 ± 0.4 Å and 6.0 ± 0.6 Å. The green lines form a perpendicular set of axes, and the purple line is rotated 7.5º with respect to one of the green lines. The angular distortion varies between different domains.

The wide bias dI/dV spectrum in Fig. 4a of the main text (from -1 eV to +1 eV) was obtained using tunneling parameters I = 500 pA and $V_s$ = -1 V, where the sample bias $V_s$ is the negative of the voltage applied to the tip. The smaller range dI/dV spectra in Fig. 4b-c of the main text were obtained using tunneling parameters I = 100 pA and $V_s$ = 0.3 V. Since sequential dI/dV spectra are often acquired on constant current isosurfaces, caution is required when comparing dI/dV intensities at different locations.

The size of the bulk gap was extracted through STS curves by performing linear fits on the two sides and the bottom of the gap in log(dI/dV) (see Fig. S6a). The two sides of the gap were each fitted with an $R^2$ > 0.95 line, and then an interval sandwiched between the two linear fits (i.e. the bottom of the gap) was fitted to a third line. We defined the bulk gap width $E_g$ to be the difference in energy between (1) the intersection of the left side linear fit and the bottom fit and (2) the intersection of the right side linear fit and the bottom fit. This is similar to a gap-finding algorithm described in Ref. 27. Because of the inhomogeneity in the samples, the gap size and energy position varies between different WTe$_2$ domains. Fig. S6b shows a histogram of the gap widths $E_g$ determined from 115 STS curves using the method depicted in Fig. S6a. dI/dV on each 1T'-WTe$_2$ domain was acquired at a random location in the bulk. We define the bulk to be ~5 nm away from an edge (and far from surface adsorbates) because the V-shaped conductance can extend up to 4 nm away from an edge (there is some variation in the edge extent). At some spots on the surface, the gap width cannot be determined with this method because there is a peak inside the bulk gap (likely due to defect states). The mean gap size is 56 meV with a standard deviation of 14 meV. The error in determining the mean gap size is at most 10 meV. Due to the bulk gap's asymmetry around the Fermi energy, the gap cannot be explained purely by phonon-assisted inelastic tunneling.[33] The measured gap sizes are plotted with respect to the domain area in Fig. S6c. No correlation between the domain area and the gap size was found, indicating that quantum confinement does not explain the gap and has no discernable effect on the electronic structure of the different domains.

The existence of the edge state is not affected by edge irregularities and roughness, as expected for a topologically protected non-trivial edge state. Fig. S7 shows an example of a dI/dV map of a rough edge. We observe that the edge state runs continuously all along a domain edge, even though the fine details of the V-shaped spectra are slightly different from point to point. It is unlikely for a topologically trivial edge state to be so continuous on edges with different terminations and disorder, and so this observation can be taken as additional evidence that we are observing non-trivial edge states[30].

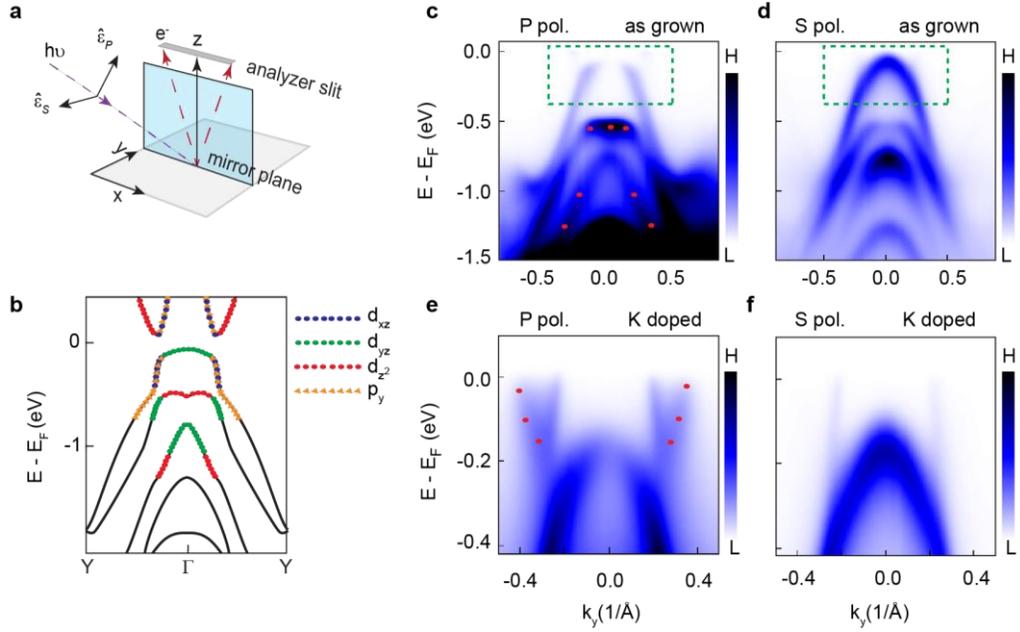

**Fig. S1. Polarization dependent ARPES.** (a) Experiment setup for polarization dependent ARPES measurement. (b) Calculated contribution (reproduced from (20)) from different orbitals for the valence and conduction bands of 1T'-WTe$_2$. (c, d) ARPES data along ΓY direction taken with P-polarized and S-polarized light, respectively. (e, f) Polarization dependent ARPES data along ΓY direction measured on a surface K-doped samples to clearly observe the conduction band only within the enlarged area of the blue boxes in (c) and (d). The red dots in (c) and (e) are added as a guide to the eye to mark orbitals with out-of-plane character. The K adatoms not only transfer charge to the film underneath, but also create extra scattering and thus broaden the FWHM of peaks. Broadened valence band and conduction band contribute a relatively high residual spectral weight between them. In the case of (e), due to the complete suppression of the half of the conduction band from the orbital selection through a polarization dependent matrix element, residual spectral weight between the conduction band and valence band makes a false impression that they were connected.

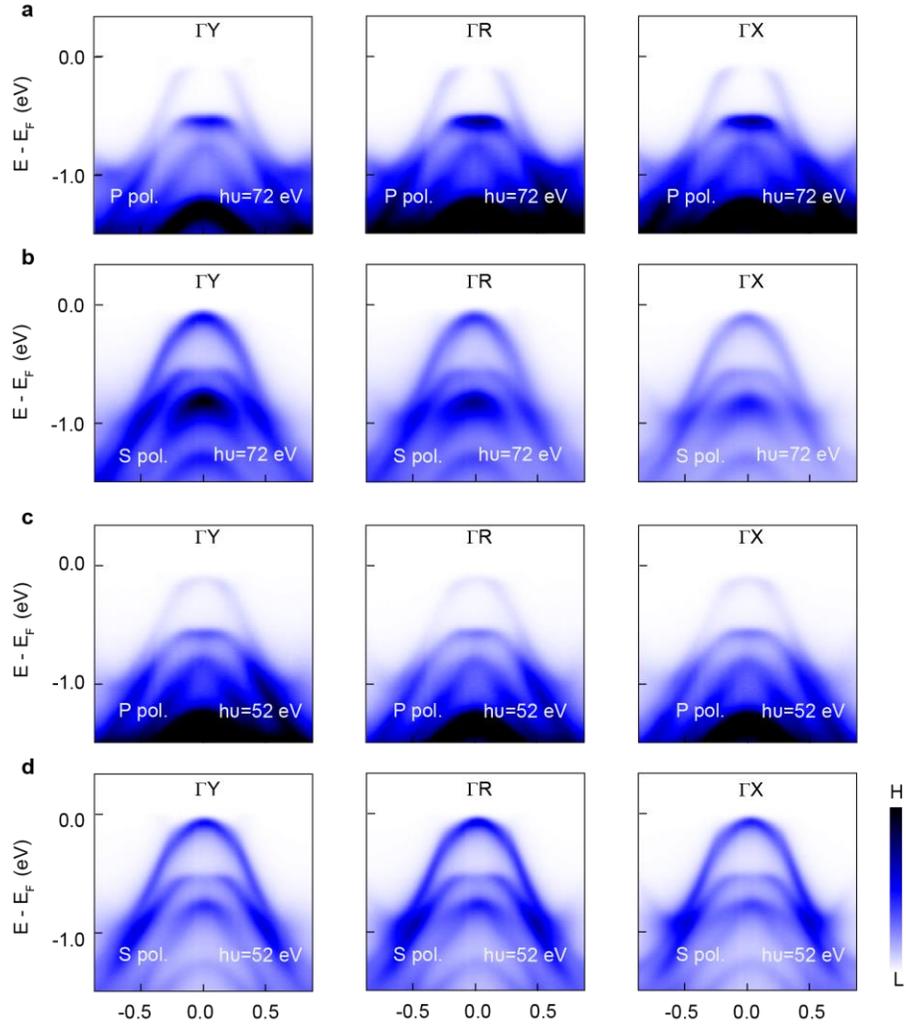

**Fig. S2.** ARPES spectra along high symmetry directions taken with different azimuth angle, polarizations and photon energies.

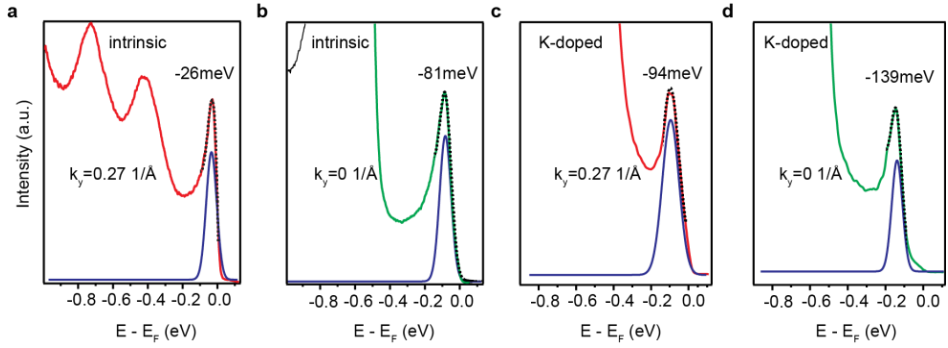

**Fig. S3. Details of the EDC fitting.** (a, b) EDCs corresponding to the conduction band bottom and the valence band top, respectively, for the samples without surface electron doping. The blue curve is the Gaussian peak, and the black dotted lines are the final fitting result with Shirley background added. (c, d) The same as (a, b) but taken for Potassium surface doped samples.

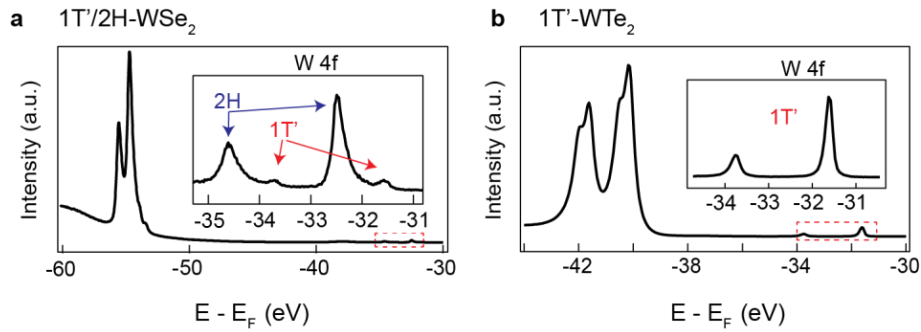

**Fig. S4.** Comparison of core level spectra of (a) mixed phase $WSe_2$ and (b) pure 1T' phase $WTe_2$. In the case of $WSe_2$, W 4$f$ peak splits into two sets due to the coexisting 1T' and 2H phase, while only one set of peaks exist in 1T'-$WTe_2$.

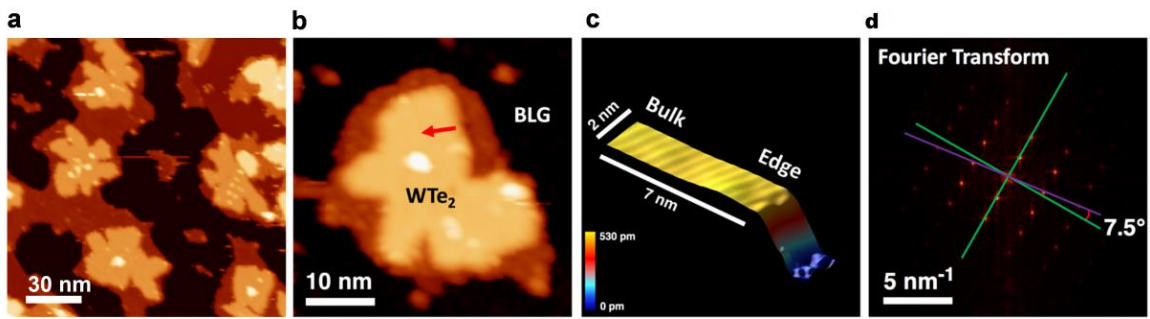

**Fig. S5. STM characterization of 1T'-WTe$_2$**. (a) STM topographic image of 1T'-WTe$_2$ film on BLG. The growth process was halted before reaching to continuous film in a deliberate effort to maximize the number of edges. The black area is bare graphene substrate. The dark brown area is a contamination layer which sits on the bare graphene and tends to be amorphous and somewhat mobile. The light orange area is monolayer WTe$_2$. There is a SiC step near the top right corner which lightens the relative colors of the graphene, contamination, and WTe$_2$ in that region. (b) STM topographic image of a single domain. The red arrow is the line along which the spectroscopy in Fig. 4c of the main text was acquired. (c) A 2 nm x 10 nm topographic image of a typical 1T'-WTe$_2$ edge. Horizontal stripes separated by ~6 Å can be seen on the 1T'-WTe$_2$. (d) Fourier transform of Fig. 2c of the main text. The green lines form a perpendicular set of axes. The purple line is rotated 7.5º with respect to one of the green lines.

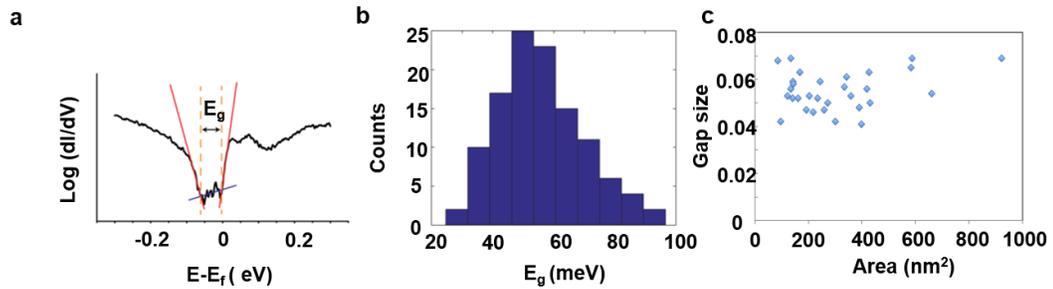

**Fig. S6. Determining the bulk gap size from STS.** (a) The bulk gap size is determined by linear fits to the two sides and the bottom of the gap. The gap width E$_g$ is the difference in the energies of the intersection points (similar to Ref. 27). (b) Histogram of gap sizes. The mean is 56 meV, and the standard deviation is 14 meV. (c) Size of the bulk gap as a function of domain size as measured by STS.

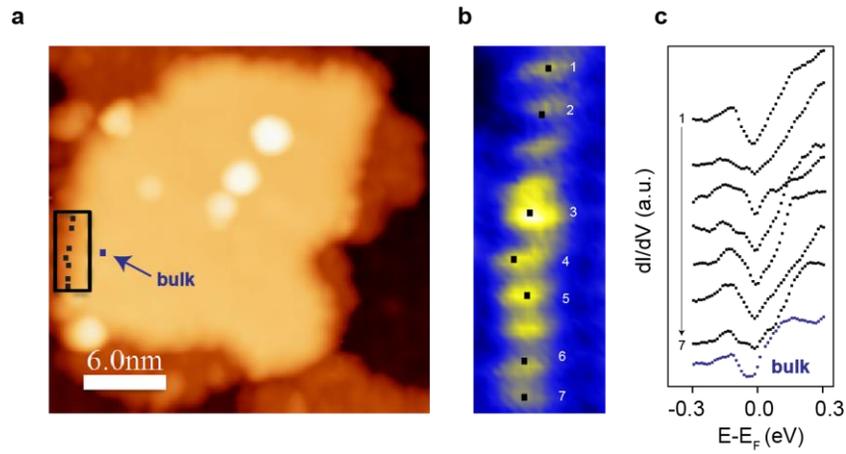

**Fig. S7. Continuous edge states on irregularly shaped edges** (a) STM image of a monolayer 1T'-WTe$_2$ domain. (b) dI/dV map at the edge of WTe$_2$ domain shown in (a) within the black rectangular box. (c) Corresponding dI/dV curves labeled in (b) except for the blue curve which is a bulk reference curve obtained at the point marked with a blue dot in (a).